\documentclass[preprint]{aastex}
\usepackage{psfig}

\newcommand{\et}{et~al.\ }
\newcommand{\Ha}{H$\alpha$}

\newcommand{\Msun}{\ensuremath{M_\odot}}

\shorttitle{X-rays from NGC 4395}
\shortauthors{MORAN ET AL.}

\begin{document}

\title{Extreme X-ray Behavior in the Low-Luminosity Active Nucleus of NGC 4395}

\author{Edward C.\ Moran,\altaffilmark{1,2}
        Michael Eracleous,\altaffilmark{3}
        Karen M.\ Leighly,\altaffilmark{4}
        George Chartas,\altaffilmark{3}
        Alexei V.\ Filippenko,\altaffilmark{2}
        Luis C.\ Ho,\altaffilmark{5} and 
        Philip R.\ Blanco\altaffilmark{6}}

\altaffiltext{1}{Chandra Fellow.}

\altaffiltext{2}{Department of Astronomy, University of California,
                601 Campbell Hall, Berkeley, CA 94720-3411.}

\altaffiltext{3}{Department of Astronomy and Astrophysics, The Pennsylvania
                 State University, 525 Davey Laboratory, University Park,
                 PA 16802.}

\altaffiltext{4}{Department of Physics and Astronomy, The University of
                 Oklahoma, 440 West Brooks Street, Norman, OK 73019.}

\altaffiltext{5}{The Observatories of the Carnegie Institution of Washington,
                 813 Santa Barbara Street, Pasadena, CA 91101-1292.}

\altaffiltext{6}{Center for Astrophysics and Space Sciences, University of
                 California, San Diego, 9500 Gilman Drive, La Jolla, CA
                 92093-0111.}

\begin{abstract}

We present the results of a 17 ks {\it Chandra\/} observation of the nearby
dwarf spiral galaxy NGC~4395, focusing on the X-ray properties of the
moderate-mass black hole that resides in its nucleus.  {\it Chandra\/}
affords the first high-quality, broadband X-ray detection of this object
that is free of contamination from nearby sources in the field.  We find
that the nuclear X-ray emission is unresolved in the {\it Chandra\/} image
and confirm the rapid, large-amplitude X-ray variability reported in previous
studies.  The spectrum of the nuclear source shows evidence for absorption
by an ionized medium.  There is also evidence for spectral variability over
the course of the {\it Chandra\/} observation, although contrary to prior
reports, it appears to be uncorrelated with fluctuations in the hard X-ray
count rate.  It is possible that the short-term spectral variability results
from column density changes in the ionized absorber.  By far the most unusual
high-energy property of NGC~4395 is the shape of its spectrum above 1~keV.
The {\it Chandra\/} data indicate a power-law photon index of $\Gamma \approx
0.6$, which is much flatter than the $\Gamma \approx 1.8$ X-ray spectra
typical of active galactic nuclei {\it and\/} the slope of the nuclear X-ray
spectrum measured from an earlier {\sl ASCA\/} observation of NGC~4395.
This extreme flatness and dramatic long-term variability of the X-ray
spectrum are unprecedented among active galactic nuclei.  A variety of
possibilities for the origin of the flat continuum slope are considered;
none provides a fully satisfactory explanation.
\end{abstract}


\newpage
\section{Introduction}

The optical/ultraviolet spectrum of the nucleus of NGC~4395, a nearby dwarf
spiral galaxy, exhibits many of the features that characterize the spectra
of luminous type~1 Seyfert galaxies and quasars (Filippenko \& Sargent 1989;
Filippenko, Ho, \& Sargent 1993).  In addition to a featureless continuum
and strong forbidden emission lines, some of which indicate a very high
degree of ionization, NGC~4395 displays broad permitted emission lines with
velocity widths of several thousand km~s$^{-1}$.  Additional evidence from
across the electromagnetic spectrum (Moran et al.\ 1999; Lira et al.\ 1999;
Kraemer et al.\ 1999; Iwasawa et al.\ 2000; Wrobel, Fassnacht, \& Ho 2001)
supports the hypothesis that NGC~4395 is powered by accretion onto a black
hole, similar to other active galactic nuclei (AGNs).

In a number of respects, however, the AGN in NGC~4395 is a prominent
outlier.  Most notably, with an absolute $B$ magnitude of $-11$ and an
\Ha\ luminosity of a few times $10^{38}$ ergs~s$^{-1}$, it is the
optically least luminous broad-line AGN currently known.  Based on
stellar velocity dispersion measurements and other arguments
(Filippenko \& Ho 2002), the mass of the black hole in NGC~4395 is
constrained to be at most $\sim 8 \times 10^5\, \Msun$, and probably
less than $\sim 10^5\, \Msun$ --- significantly smaller than the black
holes found in the nuclei of other galaxies, both active and
quiescent.  Possibly related to its low black-hole mass, the AGN in
NGC~4395 is located in a nearly bulgeless dwarf galaxy; in contrast,
other galaxies known to contain nuclear black holes are more massive
and have well-developed bulges.  NGC~4395 displays some exceptional
emission characteristics as well.  For example, it is one of the most
X-ray--variable AGNs known (Iwasawa et al.\ 2000), and its
radio--to--X-ray spectral energy distribution differs markedly from
those of other types of AGNs, perhaps indicating that there is
something special about the accretion processes in NGC~4395.  X-ray
emission is directly linked to the accretion processes in AGNs. In
this paper, we present the results of a high spatial resolution, broadband
{\it Chandra\/} observation of the central X-ray source in NGC~4395,
providing new insights into the nuclear activity in this unusual
object.  We adopt a distance of 4.1 Mpc for NGC~4395, based on {\it
Hubble Space Telescope\/} observations of the brightest asymptotic
giant branch stars in the galaxy (Minitti et al.\ 2002).

\section{The {\it Chandra\/} Observation}

NGC 4395 has been the target of pointed X-ray observations in the 0.1--2.4
keV band with {\sl ROSAT\/} (Moran \et 1999; Lira \et 1999) and in the
0.6--10 keV band with {\sl ASCA\/} (Iwasawa \et 2000).  However, because
of the modest spectral coverage and collecting area of {\sl ROSAT\/} and
the poor angular resolution of {\sl ASCA}, a complete picture of the
high-energy properties of NGC 4395 is lacking.  {\it Chandra\/} provides
the first broadband, high-resolution X-ray view of this object.

NGC~4395 was observed with {\it Chandra\/} on 20 June 2000 (UT) for 17,187~s
with the ACIS-S instrument.  The S3 chip was the only active CCD; a 1/2
subarray was read out to reduce the possibility of photon pile-up, which
yielded a frame time of 1.54~s.  The nucleus of the galaxy was located at
the instrument aim point.  We reprocessed the data with the CIAO software,
version 2.1.2, in order to remove the 0\farcs5 pixel randomization introduced
during the standard processing of the data.  The screened data consist of
events with grades 0, 2, 3, 4, and 6; no background flares occurred during
the observation.  In the following sections, we describe the X-ray imaging,
variability, and spectroscopy results obtained with {\it Chandra}.

\section{X-ray Imaging}

The {\sl ROSAT\/} PSPC image of NGC~4395 revealed that there are at least
five X-ray sources within the central 2\farcm8 of the galaxy (Moran \et
1999), one of which is significantly brighter than the nucleus at low
X-ray energies.  Thus, in the images obtained with {\sl ASCA\/} (half-power
diameter $\approx 3'$), the nuclear source was significantly contaminated
by emission from the other nearby sources, particularly below 2 keV (Iwasawa
\et 2000).  As Figure~1 illustrates, the nucleus (source A) is completely
isolated in the $\sim 1''$ resolution {\it Chandra\/} image.\footnote{As
a result of the subarray employed in our 17~ks exposure, a bright nearby
source (source E) was located outside the field of view.  It was, however,
included within the field of a short 1260~s ACIS-S observation of NGC~4395
(see Ho et al.\ 2001), which we have collected from the {\it Chandra\/}
data archive.  Figure~1 is a superposition of the two datasets; the true
strength of source~E is therefore not represented in the figure.}
The high angular resolution of {\it Chandra\/} also allows us to measure
the position and spatial extent of the nuclear X-ray emission in NGC~4395
with unprecedented accuracy.  The centroid of the nuclear source is
$\alpha$(2000) = $12^{\rm h} 25^{\rm m} 48.\!\!^{\rm s}$84,
$\delta$(2000) = $33^{\circ} 32' 48.\!\!{''}9$,
just 0\farcs4 from the measured position of the VLBA radio source (Wrobel \et
2001) that is coincident with the optical nucleus.  The X-ray source offset
is well within the $\sim$~1\farcs5 {\it Chandra\/} position uncertainty,
which should lay to rest any lingering doubt about the association between
the central X-ray emission and the nucleus of NGC~4395.  The radial profile
of the nuclear X-ray source indicates that it is unresolved; the fractional
encircled energy within a $2''$ radius is about 97\%.

A $5''$ radius aperture was used to extract light curves and the spectrum
of the nucleus.  The background was measured in a concentric annulus with inner
and outer radii of $20''$ and $40''$, respectively.  The background annulus
lies between the nucleus and the nearest off-nuclear source (source B
in Fig.~1).  A net total of 2374 counts were detected in the 0.3--10.0 keV
energy range.  Note that this is a factor of 20 greater than the number of
counts detected in the {\sl ROSAT\/} PSPC exposure of NGC~4395 (Moran \et
1999), which had almost exactly the same duration.

\section{X-ray Variability}

Previous X-ray observations of NGC~4395 have established the variable nature
of its nucleus.  In the {\sl ROSAT\/} band, the nuclear source varied by a
factor of three over the several days spanning the PSPC observation (Moran
et al.\ 1999).  Rapid variability (with doubling times on the order of 100~s)
was witnessed during the {\sl ASCA\/} observation at energies greater than
2~keV, where the nucleus was more clearly isolated from other sources in
the field (Iwasawa et al.\ 2000).  Our high signal-to-noise ratio ($S/N$)
{\it Chandra\/} data are free of contamination over the broad 0.3--10 keV
range, and unlike the {\sl ROSAT\/} and {\sl ASCA\/} observations, the
{\it Chandra\/} exposure is continuous.

The full-band light curve for the nucleus of NGC 4395 is displayed in
Figure~2.  A bin size of 77~s (= 50 frame times) has been employed.
As the figure indicates, the source was highly variable during the 17~ks
{\it Chandra\/} observation; the count rate fluctuated by an order of
magnitude during the exposure, with dramatic changes (factors of 2--3)
occurring over very short periods.  While some of the strong flares and
dips in the light curve [e.g., the flickering in the $t \approx (0.7 - 1.3)
\times 10^4$~s range] appear to be time-resolved at this binning, there
are a few instances where the count rate increased and then decreased
(or vice versa) by a factor of $\sim 2$ over the span of a single 77~s
bin.  This time scale constrains the size of the emitting region to be
less than 77 light-seconds ($\sim 2 \times 10^{12}$ cm).

To investigate whether the variability of the source has any spectral
dependence, we have constructed light curves in different energy bands:
hard (2--10 keV; ``H"), medium (1--2 keV; ``M"), and soft (0.3--1.2 keV; ``S").
As shown in the top two panels of Figure~3, the hard-band and full-band light
curves are very similar, both in terms of the features present and the
total intensity.  Thus, most of the detected source counts have energies
above 2~keV.  However, the light curves in the medium and soft energy
bands (third and fourth panels of Fig.~3) provide important additional
information about the variability properties of NGC~4395.  While the
count rates in all three bands are elevated in the $t \approx 7000$--9000~s
range, there are also clear differences between the H, M, and S light
curves. Of particular interest are the dip in the 2--10 keV count rate at
$t \approx 5500$~s, which is absent in the soft and medium energy bands,
and the strong increase in the soft-band count rate at $t \approx$
11,500~s, which is not apparent above 2~keV.  These differences result
in significant spikes in the X-ray ``colors'' of the source (i.e., the
S/H and M/H count-rate ratios), which are shown in the bottom two panels
of Figure~3.  Sharp color variations are evident at other times as well,
but in general they do not appear to correspond directly with changes in
the 2--10 keV count rate.  As discussed in the following sections, these
light curve properties play an important role in the interpretation of
spectral models for the nuclear X-ray source. 

\section{X-ray Spectroscopy}

X-ray spectroscopy provides vital clues about the emission and absorption
components present in the nucleus of NGC~4395, as well as its intrinsic
luminosity.  Interesting spectral properties of this source have already
been revealed in previous X-ray observations.  For example, even though
the $S/N$ of the {\sl ROSAT\/} PSPC spectrum was poor, we were unable to
obtain a good fit using a power-law model with absorption by a Galactic
column of neutral material (Moran \et 1999), which in the {\sl ROSAT\/}
band describes the X-ray spectra of most broad-line AGNs (e.g., Walter
\& Fink 1993).  The absence of strong spectral variability corresponding
to the factor-of-three change in the source flux suggested to us that a
single emission component dominates below 2 keV; we further speculated
that an ionized (or ``warm'') absorber (Halpern 1984) might be responsible
for the spectral complexity.  The subsequent {\sl ASCA\/} observation of
NGC~4395 appeared to indicate an excess of soft X-ray emission during
periods when the 2--10 keV count rate was high.  To account for this
behavior, Iwasawa \et (2000) adopted a model consisting of a power-law
continuum and two components of ionized absorption, one that is constant
and one that is variable.  In this model, the ionization parameter of
the latter component increases when the source is in an active state,
resulting in a lower opacity in the 1--2 keV range and excess observed flux
in this band.  Given the possibility that the {\sl ASCA\/} spectrum was
contaminated below 2 keV by emission from nearby sources in the field, this
scenario should be reexamined with the aid of {\it Chandra}'s resolution.

The ACIS count rate is high enough (at times) to raise concerns about photon
pile-up (the occurrence of two or more events in a pixel within one frame
time), which could affect the shape of the observed spectrum.  The first
indication that pile-up is probably not significant in our observation of
NGC~4395 is the fact that the strong edge in the mirror response near 2.1
keV is present in the spectrum; this feature is usually smeared out when
pile-up is severe.  Simulations with the LYNX tool (Chartas \et 2000;
Eracleous \et 2001) confirm that pile-up effects are minimal: the pile-up
fraction is just 3\% during periods of relative quiescence [e.g., the $t
= (1.2-1.7) \times 10^4$~s range], and during most active phase (i.e., the
$t$ = 7000--9000~s range), the pile-up fraction is still only $\sim$~10\%.

The bottom panels of Figure 3 indicate that the spectrum of NGC~4395
fluctuates somewhat with time.  Although it would appear that spectral
variations and features in the hard X-ray light curve are uncorrelated, the
issue needs to be investigated further, since any dependence of the spectral
properties on count rate would require us to model the high-state and low-state
spectra separately.  Unfortunately, given the degree of structure in
the light curve shown in Figure~2, it is difficult to divide the data into
temporally distinct active and quiescent phases, as Iwasawa \et (2000)
did for the {\sl ASCA\/} observation.  However, the source was clearly
more active at certain times during the {\it Chandra\/} exposure than
at others; we therefore extracted an ``active'' spectrum using data in
the $t = 6900$--10,400~s range and
a ``quiescent'' spectrum using data in the $t = 1200$--2400, 5000--6000,
and 12,000--16,800~s ranges.  These spectra contain 28\% and 20\% of the
total counts, respectively.  As an alternative approach, we have derived
the distribution of count rates in the light curve, which is displayed
in Figure~4.  Based on this histogram, we have compiled high-state and
low-state spectra of the source by collecting counts in periods represented
by the highest 6 bins ($> 0.24$ count s$^{-1}$) and the lowest 5 bins ($< 0.15$
count s$^{-1}$) separately.  These spectra contain $\sim 25$\% and 40\%
of the detected counts, respectively.  In Figure~5 we have plotted the
ratio of the high-state and low-state spectra derived using both of the methods
described here.  Despite the flux difference of a factor of $\sim 3$
between the two spectra in each case, spectral variations are only
marginally significant.  This is contrary to the results of the {\sl
ASCA\/} observation reported by Iwasawa \et (2000).  We note that the
shape of the spectral ratio shown in Figure~5b, derived using the
count-rate histogram in Figure~4, {\it does\/} bear some resemblance
to a similar plot presented by Iwasawa \et (2000).  However, the
1--2 keV excess we observe relative to the 4--10 keV ratio (40\%)
is far less than the factor of 2--3 excess indicated in the {\sl ASCA\/}
data.  For the purposes of modeling the ACIS-S spectrum, we have no
evidence that the spectrum changes significantly in direct response to
count-rate fluctuations.  Thus, in the analyses that follow, we model
the spectrum derived from the full {\it Chandra\/} data set.

For model fitting, the {\it Chandra\/} spectrum of NGC~4395 was grouped into
bins containing a minimum of 25 counts.  Because of uncertainties in the
ACIS-S response at the lowest and highest energies, we have ignored channels
below 0.5 keV and above 9 keV.  A power-law model with absorption by neutral
material provides a poor fit to the {\it Chandra\/} spectrum over the entire
0.5--9 keV range. It does, however, provide an excellent fit to the data
above $\sim 1.2$ keV.  The best-fit photon index and absorption column density
(with 90\% confidence ranges for two parameters of interest) are
$\Gamma =0.61^{\scriptscriptstyle +0.25}\!\!\!\!\!\!\!\!\!\!\!
             _{\scriptscriptstyle -0.20}$ and
$N_{\rm H} = 1.2^{\scriptscriptstyle +0.4}\!\!\!\!\!\!\!\!\!
                _{\scriptscriptstyle -0.3} \times 10^{22}$ cm$^{-2}$; the
fit is displayed in Figure~6.  An unresolved Fe~K$\alpha$ emission line at
6.4 keV is very marginally detected with an equivalent width of $99 \pm 95$
eV (90\% confidence).  The improvement to the fit obtained with the inclusion
of this component ($\Delta \chi^2 = 2.98$ for one additional free parameter)
is significant at the 95\% confidence level.
Note that the power-law index we obtain is substantially lower than the
value of $\Gamma = 1.72$ obtained by Iwasawa et al.\ (2000) from
their analysis of the {\sl ASCA\/} observation.  We return to this point
in \S~6.4.

One means of explaining the apparent excess flux below $\sim 1$ keV would
be to invoke a second emission component that dominates at low energies.
We obtain an excellent fit over the full energy range if this additional
component is an optically thin thermal plasma (either a Raymond-Smith or
``MEKAL'' component) with a temperature of $\sim 0.1$ keV.  However, as
Figure~3 indicates, the source is quite variable in the 0.3--1.2 keV band.
The time scale of the variability, the temperature of the thermal component,
and its implied luminosity combine to suggest that such a plasma would be
optically thick (Elvis et al.\ 1991), which rules out composite models of
this sort.  We also tried composite models where the second component is
either a blackbody or another (steeper) power law; such components could
plausibly be associated with a compact emission region near the black hole.
These models also provide good fits (although the residuals are more
pronounced below 1 keV).  They do not, however, easily explain why the
soft-band and hard-band light curves are similar in some respects and
dissimilar in others.  The most likely scenario, in our minds, is that a
single emission component dominates over the entire 0.3--10 keV band.  

To investigate this possibility, we have refitted the {\it Chandra\/}
spectrum above 2.2 keV with a single power-law model, assuming only
the Galactic neutral hydrogen column density ($1.4 \times 10^{20}$
cm$^{-2}$; Murphy \et 1996).  The residuals associated with this fit,
displayed in Figure~7, exhibit several strong dips in the soft X-ray
band.  The deepest feature at $\sim$~0.8 keV is consistent with the
absorption edges of \ion{O}{7} (0.74 keV) and \ion{O}{8} (0.85 keV), which
suggests that an ionized absorber (Halpern 1984; Reynolds 1997) is
present in NGC~4395.  Indeed, with the addition of a warm absorber
to the model (i.e., the ``absori'' component in the XSPEC spectral
analysis package), excellent fits to the {\it Chandra\/} spectrum over
the full energy range are possible.  Assuming a fixed temperature of
$1 \times 10^5$~K for the absorber, we obtain an acceptable fit
($\chi^2 = 67.6$ for 79 degrees of freedom) with the following model
parameters: a power-law photon index $\Gamma = 0.54$, a column
density of the ionized medium $N_{\rm warm} = 1.37 \times 10^{22}$
cm$^{-2}$, and an ionization parameter $\xi = 6.4$ erg cm s$^{-1}$.
A slightly better fit ($\chi^2 = 63.7$) is obtained assuming $T = 5
\times 10^4$~K; the fit parameters (with uncertainties corresponding
to $\Delta \chi^2 = 2.71$) are
$\Gamma = 0.56^{\scriptscriptstyle +0.13}\!\!\!\!\!\!\!\!\!\!\!
             _{\scriptscriptstyle -0.13}$,
$N_{\rm warm} = 1.49^{\scriptscriptstyle +0.31}\!\!\!\!\!\!\!\!\!\!\!
             _{\scriptscriptstyle -0.29} \times 10^{22}$ cm$^{-2}$,
and
$\xi = 9.3^{\scriptscriptstyle +4.1}\!\!\!\!\!\!\!\!\!
                _{\scriptscriptstyle -3.1}$ erg cm s$^{-1}$.
This best-fit model is displayed in Figure~8.  In general, as the assumed
temperature is increased, the model ionization parameter decreases and
the fit below 1.3 keV worsens.

Although not required by the data, we have also fitted a model
consisting of two components of ionized absorption similar to the
multizone warm absorber model used by Iwasawa et al.\ (2000).  The
temperatures of the two components are assumed to be $5 \times 10^4$~K
and $1 \times 10^6$~K.  Fixing the photon index at a value of 0.56,
the improvement to the fit obtained with this model ($\Delta \chi^2 =
2.54$ for one additional free parameter) is significant at the 92\%
confidence level.  The column density and ionization parameter of the
``cool'' warm absorber component are $N_{\rm warm} = 1.0 \times
10^{22}$ cm$^{-2}$ and $\xi = 4.6$ erg cm s$^{-1}$.  For the ``hot''
component, we find $N_{\rm warm} = 1.4 \times 10^{22}$ cm$^{-2}$ and
$\xi = 42$ erg cm s$^{-1}$.  Given the flexibility of the model and
the quality of the data, it is not too surprising that these
parameters are not very well constrained.  Additional observations
with higher $S/N$ are needed to give us a better handle on the
physical state of the ionized gas in NGC~4395, particularly if the
absorption varies with time, as the difference between our results and
those obtained with {\sl ASCA\/} might suggest.  The simple fact that
warm absorber models afford such good fits is important, since it
supports the notion that the broadband X-ray spectrum of NGC~4395 is
dominated by a single emission component.  The extremely flat slope of
this component, however, is very surprising; in the next section we
consider a variety of possibilities for its origin.

\section{Implications for the Nature of the Nucleus of NGC 4395}

\subsection{Broadband Luminosity and Variability of the Source}

Using the high angular resolution of {\it Chandra}, we have established that
the central X-ray source in NGC~4395 is coincident with the galaxy's optical
and radio nucleus.  The X-ray emission is spatially unresolved and highly
variable, leaving no doubt that it is associated with the dwarf AGN in the
galaxy.  Provided that a single emission component dominates in the
{\it Chandra\/}
band, the power-law fit to the ACIS-S spectrum above 1.2 keV suggests an
absorption-corrected, time-averaged 2--10 keV luminosity of $8.0 \times
10^{39}$ ergs~s$^{-1}$ for an adopted distance of 4.1 Mpc.  This is within
25\% of the value obtained with {\sl ASCA\/} (adjusted for this distance)
by Iwasawa \et (2000).  In the 0.5--10 keV band, $L_{\rm X} = 9.0 \times
10^{39}$ ergs~s$^{-1}$.  Combining the broadband X-ray luminosity with the
radio--to--UV luminosity computed by Moran \et (1999, after revision for
the 4.1 Mpc distance), we obtain a bolometric luminosity of $L_{\rm bol} =
5.3 \times 10^{40}$ ergs~s$^{-1}$.  This luminosity corresponds to an
Eddington-limit mass of $\sim 400$ $\Msun$ for the accreting black hole.
As reported by Filippenko \& Ho (2002), the conservative upper limit for
the black-hole mass is $8 \times 10^5$ $\Msun$; the actual mass, they
argue, is likely to be between $\sim 1 \times 10^4$ $\Msun$ and $\sim 1
\times 10^5$ $\Msun$.  Using this range of values, the bolometric luminosity
suggests an Eddington ratio of $L_{\rm bol}/L_{\rm edd} = 0.004$--0.04.

The nucleus of NGC 4395 is an exceptionally variable X-ray source.  As pointed
out by Iwasawa \et (2000), such extreme variability is quite uncharacteristic
of low-luminosity AGNs (LLAGNs).  The ``excess variance'' (Nandra \et 1997;
Turner \et 1999) of the source in the 2--10 keV band is $0.35 \pm 0.03$
(using 128~s bins), almost twice that estimated by Iwasawa \et using
{\sl ASCA\/} data and far greater than the excess variances measured for
other LLAGNs (e.g., Ptak \et 1998).  However, many of the LLAGNs examined
to date are LINERs, which are likely to be accreting in an advection-dominated
mode (e.g., Lasota et al.\ 1996; Ho 1999) --- in other words, they are
supermassive black holes ($M > 10^6$ $\Msun$) with low accretion rates
($\dot{M}/\dot{M}_{\rm edd} < 10^{-3}$).  In contrast, NGC 4395 appears
to possess a smaller black hole that is accreting at a higher rate.

\subsection{Ionized Absorption and the Short-Term Spectral Variability}

The X-ray spectrum of NGC 4395 below $\sim 1.2$ keV is complex, indicating
that either a second emission component dominates at low energies or that
an ionized absorber is present.  The rapid variability of the source in
the soft X-ray band, and the fact that some of the count-rate fluctuations
observed above 2 keV occur at lower energies as well, argue for the latter
interpretation.  We obtain an excellent fit to the ACIS-S spectrum with a
model consisting of a power-law continuum, neutral absorption equivalent to
the Galactic column, and a single-temperature component of ionized absorption.
If a simple, constant warm absorber were present, however, we would expect
the light curve below 1 keV to closely resemble the hard-band light curve,
which it does not.  On the other hand, the more complex multizone warm
absorber model proposed by Iwasawa \et (2000) cannot explain our results
either.  In that scenario, the opacity of a variable absorption component
changes in response to intensity fluctuations of the source, which leads to
excess soft X-ray emission when the source is in a high-flux state.  The
{\it Chandra\/} light curves in Figure~3 and the spectral ratio plots shown
in Figure~5 do not indicate this sort of behavior.  In general, excess soft
X-ray emission does not accompany an increase in the hard-band count rate,
and in at least one instance, we observe a dramatic softening of the spectrum
corresponding to a sharp {\it decrease\/} in the hard X-ray flux.

A possible explanation for the short-term spectral variability we observe
is that the absorbing medium in NGC~4395, rather than being physically
variable, is {\it geometrically\/} variable or clumpy.  The spatial
scale of the clumpiness and the transverse velocity of the medium
could, in the right combination, give rise to rapid column density
variations along the line of sight, and thus, to changes in the
observed soft X-ray flux that are partially independent of the
continuum variability above 2 keV.  Soft-band and hard-band light
curves that are similar in some respects, but not in others, would be
a natural consequence of this scenario.  As a test, we have used the
PIMMS software to compute the ACIS-S count rates that would be
observed in various bands if the $\Gamma = 0.6$ power law were
absorbed only by a Galactic column of neutral material.  Relative to
the 2--10 keV band, we would expect 0.65 and 0.73 times as many counts
in the 0.3--1.2 keV and 1--2 keV bands, respectively.  If a clumpy
absorbing medium in NGC~4395 is responsible for differences in the
soft-band and hard-band light curves, the X-ray colors of the source
defined in Figure~3 should not exceed these amounts.  Referring to the
figure, we see that the maximum value of Color~1 (the [0.3--1.2
keV]/[2--10 keV] ratio) is $\sim 0.4$ and the maximum value of Color~2
(the [1--2 keV]/[2--10 keV] ratio) is $\sim 0.7$, consistent with our
hypothesis.  To further investigate the role of column density
variations, we have used the Color~1 light curve to predict the values
of Color~2 for the entire {\it Chandra\/} observation.  This was done
by adjusting the column density of the ionized medium in the best-fit
single warm absorber model (that shown in Fig.~8) until the observed
values of Color~1 were obtained.  The same model, with the modified
column density, was then used to derive Color~2.  The range of column
densities implied is $\sim 1 \times 10^{22}$ to $\sim 4 \times 10^{22}$
cm$^{-2}$.  As Figure~9
illustrates, the agreement between the observed and predicted values
of Color~2 is quite good.  Unfortunately, the uncertainties are rather
substantial, and it is too soon to conclude that column density
variations alone are responsible for the X-ray color changes observed
in NGC~4395.  These results, however, are encouraging.

\subsection{Origin of the Hard X-ray Continuum}

Although a power law provides a perfectly adequate description of the
hard-band spectrum, the spectral index we measure ($\Gamma \approx
0.6$) is substantially flatter than the typical slopes of AGN spectra,
which are generally in the $\Gamma \approx$ 1.7--1.9 range (e.g.,
Nandra \& Pounds 1994). To our knowledge, an intrinsic X-ray spectrum
as flat as that of NGC~4395 has never been observed in an AGN.  Thus, it
would be very worthwhile to investigate the reason the spectrum of this
source is so hard.  First, we consider the possibility that the nuclear
spectrum of NGC~4395 is intrinsically flat.  It is generally believed
that hard X-rays in AGNs are produced by low-energy photons (perhaps
from an accretion disk) that are inverse-Compton scattered by a corona
of hot ($\sim 100\,\rm keV$) thermal electrons with fairly low optical
depth.  This mechanism, however, cannot yield a spectrum with a photon
index less than 1; a flatter spectrum would imply that scattered photons
at higher energies have acquired more energy per scattering than those
at lower energies, which is impossible.  If the scattering optical depth
is large, a flat spectrum in the band corresponding to the Wien peak of
the electron energy distribution is possible.  This circumstance, however,
implies that the photons undergo many scatterings, which would smear out
the rapid variability we observe (unless the emitting region is {\it much\/}
smaller than the limit set by the variability time scale---see \S~4).
Comptonization by a nonthermal
population of electrons cannot be ruled out, although an extreme electron
energy spectrum would be required in this scenario.  For a power-law
distribution of electron energies with index $p$, the energy index of
inverse-Compton radiation $\alpha$ ($= \Gamma - 1$) is $(p - 1)/2$ (e.g.,
Tucker 1975).  Thus, for $\Gamma = 0.6$, a $p = 0.2$ electron spectrum
is required --- vastly different from the $p \approx 2.2$ spectrum inferred
for the electrons that produce the compact synchrotron radio emission
in the nucleus of NGC~4395 (Ho \& Ulvestad 2001).  Thermal bremsstrahlung
emission from a hot, optically thin plasma produces a very flat spectrum
at X-ray energies below the characteristic temperature $kT$ of the plasma.
However, bremsstrahlung models (even those with very high temperatures)
have too much curvature to provide good fits to the spectrum of NGC~4395
above 2 keV.  Furthermore, the small size of the source set by the
variability and the implied emissivity suggest a large optical depth in
the plasma, which firmly rules out bremsstrahlung emission (e.g., Elvis
et al.\ 1991).

Alternatively, perhaps the intrinsic spectrum of NGC~4395 is steep,
but modified in such a way that a flat spectrum is observed.  Compton
reflection (e.g., Matt et al.\ 1996) is an obvious candidate for this
effect; reprocessing by ``cold'' material can result in a very flat
observed spectrum if the primary steep-spectrum emission from the source
is obscured.  Two factors lead us to conclude that reflection does not
provide a viable solution.  First, reflection-dominated spectra should
be accompanied by a strong (1--2 keV equivalent width) Fe~K$\alpha$
emission line at 6.4 keV, but there is no evidence for such a strong line
in the spectrum of NGC~4395 (Fig.~6).  A second constraint is provided
by the rapid variability of the source, which implies a maximum source
size that is only somewhat larger than the region where the primary
hard X-ray emission is expected to originate ($\sim 6 \times 10^{11}$ cm,
if $M_{\rm BH} = 10^5\, \Msun$ and the hard X-rays are produced within
about 10 Schwarzschild radii).  For reflection to dominate, obscuration
would have to cover the primary emission but not the reflector, which
seems unlikely.  Another possibility is that NGC~4395 has an intrinsically
steep spectrum that is flattened by a complex neutral absorber consisting
of multiple ``partial covering'' components.  In this scenario, a fraction
of the steep-spectrum emission is seen directly, but the majority is
observed via transmission through different columns of dense material.
The result is a flat observed spectrum that has a roughly power-law form
(e.g., Zdziarski et al.\ 2001).  Indeed, we obtain a very good fit to
the spectrum of NGC~4395 above 1.2 keV with a model consisting of a
$\Gamma = 1.7$ power law absorbed by three partial-covering components.
The absorbers in this model have fairly high columns (1--90 $\times
10^{22}$ cm$^{-2}$) and covering fractions (0.5--0.9).  Interestingly,
we find that this model provides a reasonable fit to the entire ACIS-S
spectrum.  The fit below 1.2 keV is not as good as that obtained with
a single warm absorber, but this is easily rectified with the addition
of a modest amount of ionized absorption.  The intrinsic 2--10 keV
luminosity of the source implied in this model is higher than the one we
reported above for a $\Gamma = 0.6$ power law, but only by a factor of
2.5.  We note, however, that this model predicts a strong Fe~K edge near
7~keV, which does not appear to be present in the {\it Chandra\/} spectrum.  

\subsection{Dramatic Long-Term Spectral Variability}

Of great importance is the possibility that the shape of the broadband
continuum in NGC~4395 has undergone a dramatic change during the two years
between the {\sl ASCA\/} and {\it Chandra\/} observations.  This is implied
by the difference in the power-law photon index we measure ($\Gamma \approx
0.6$) and that obtained with {\sl ASCA\/} ($\Gamma \approx 1.7$) by Iwasawa
et al.\ (2000).  Such behavior could provide important new insight into the
accretion processes not only in NGC~4395, but perhaps generally in active
galactic nuclei.  

As a first step, we have examined the spectrum of the nuclear source in
NGC~4395 during the short 1260~s {\it Chandra\/} exposure obtained by Ho
et al.\ (2001), which was carried out about two months before our 17~ks
observation.  Despite the limited $S/N$
ratio of the snapshot spectrum, an absorbed power-law model provides a good
fit above $\sim 1.2$ keV; we find that $\Gamma < 1.0$ (90\% confidence).
Thus, we need to consider the possibility that the spectrum of the nuclear
emission is always flat and appears steep in the {\sl ASCA\/} data due to
contamination by nearby sources in the field.  The large $6' \times 6'$
SIS extraction area used by Iwasawa et al.\ to investigate the spectrum
above 2 keV is approximately the same region shown here in Figure~1.  They
omitted a small region centered on source~E, but estimated that $\sim$~30\%
of the counts from E were included in the extraction.  To approximate the
{\sl ASCA\/} spectrum, we have summed the emission from the nucleus and all
of the other resolved sources in Figure~1.  The contribution of source~E,
which was present in the 1260~s exposure but not our 17~ks observation,
was determined by scaling the number of counts in the snapshot spectrum
by the exposure-time ratio and then multiplying by 0.3.  We find that in
the 0.3--10 keV band, only 62\% of the counts in the simulated {\sl ASCA\/}
spectrum are associated with the nucleus; 16\% of the counts are associated
with source~E, and the remaining 22\% are from the other resolved sources.
However, most of the contamination occurs at low X-ray energies --- a
power-law fit to the simulated {\sl ASCA\/} spectrum above 2 keV indicates
a flat spectral index similar to that which we measured for the nucleus
alone.  We conclude, therefore, that if the spectrum of the nucleus was
flat during the {\sl ASCA\/} observation, it would have appeared as such
to Iwasawa et al., implying that the slope of the broadband continuum of
NGC~4395 is indeed capable of dramatic variations.

The other significant difference between our results and those obtained with
{\sl ASCA\/} concerns the variability of the nucleus in the 1--2 keV band
and its dependence on the source count rate at higher energies.  A strong
excess 1--2 keV flux associated with increases in the 2--10 keV count rate
was seen in the {\sl ASCA\/} data (Iwasawa et al.\ 2000), but not in the
{\it Chandra\/} observation presented here.  Contamination of the soft-band
{\sl ASCA\/} spectrum is severe --- we find that over 60\% of the counts below
2 keV in the large SIS aperture employed by Iwasawa et al.\ are associated
with contaminating sources --- providing a possible explanation for the
different
results.  On the other hand, the 1--2 keV excess reported by Iwasawa et al.\
was measured in SIS spectra extracted with a smaller $3'$ diameter aperture,
which would have had a reduced level of contamination.  In view of the drastic
difference we observe in the slope of the nuclear spectrum and the possibility
of column density variations in NGC~4395 (see above), different behavior in
the 1--2 keV band during the {\sl ASCA\/} observation seems entirely plausible.
This and other issues regarding the nature of NGC~4395 should be explored with
future high-resolution X-ray observations.

\section{Summary}

The nucleus of NGC~4395 is unique among type~1 AGNs.  In addition to having
one of the lowest observed luminosities and least massive black holes, it
displays a number of extreme properties in the X-ray band as well.  As our
{\it Chandra\/} observation confirms, NGC~4395 has a spectacular X-ray light
curve that exhibits large-amplitude variations on extremely short time scales.
This object is one of the most X-ray--variable AGNs known.  It appears that
ionized gas is responsible for the bulk of the X-ray absorption in NGC~4395;
variations in the column density of this material may be the cause of the
short-term spectral variability we detect over the course of the
{\it Chandra\/} exposure.  

But the most extreme aspect of the nuclear X-ray source in NGC~4395
is the slope of its hard X-ray spectrum, which is modeled as a $\Gamma = 0.6$
power law.  This is significantly flatter than the X-ray spectral slopes
associated with more luminous type~1 AGNs ($\langle\Gamma\rangle \approx 1.8$).
Unfortunately, it remains unclear {\it why\/} the spectrum of NGC~4395 is
so flat --- Comptonization requires an extremely small, optically thick
scattering region or an electron population with very unusual energetics,
and none of the alternative possibilities we have explored
(e.g., multiple partial-covering absorption components)
provides a fully satisfactory explanation for the
flat slope.  Even more intriguing, however, are the dramatic changes in the
X-ray absorption and spectral characteristics of NGC~4395 that have apparently
occurred during the two years between the {\sl ASCA\/} and {\it Chandra\/}
observations.  Analysis of the {\sl ASCA\/} data by Iwasawa et al.\ (2000)
revealed a ``normal'' X-ray spectral slope of $\Gamma = 1.7$ for NGC~4395
and excess flux in the 1--2 keV band corresponding to increases in the 2--10
keV count rate, neither of which is evident in the {\it Chandra\/} data.
Our preliminary analysis suggests that contamination of the nuclear emission
in the {\sl ASCA\/} image by nearby sources is not responsible for these
differences.  If so, we must conclude that the nucleus of NGC~4395 undergoes
dramatic changes in the slope of its continuum and/or its absorption
properties on relatively short time scales.  Spectral variations of this
magnitude have never been observed in an AGN.  

Obviously, X-ray monitoring
of NGC~4395 with high-resolution instrumentation is crucial for a clarification
of the physical processes occurring in its nuclear region.  Ultimately, a firm
understanding of the unusual behavior in this object may provide general
insight into the physics of AGNs.

\acknowledgments

We are grateful to Andrzej Zdziarski for helpful discussions about X-ray
continuum models.
E.C.M.\ is supported by NASA through Chandra Fellowship grant PF8-10004
awarded by the Chandra X-ray Center, which is operated by the Smithsonian
Astrophysical Observatory for NASA under contract NAS 8-39073.  A.V.F.\
thanks NASA for financial support through {\it Chandra\/} grant GO0-1170A,
and the Guggenheim Foundation for a fellowship.

\clearpage
\begin{figure}
\begin{center}
\centerline{\psfig{figure=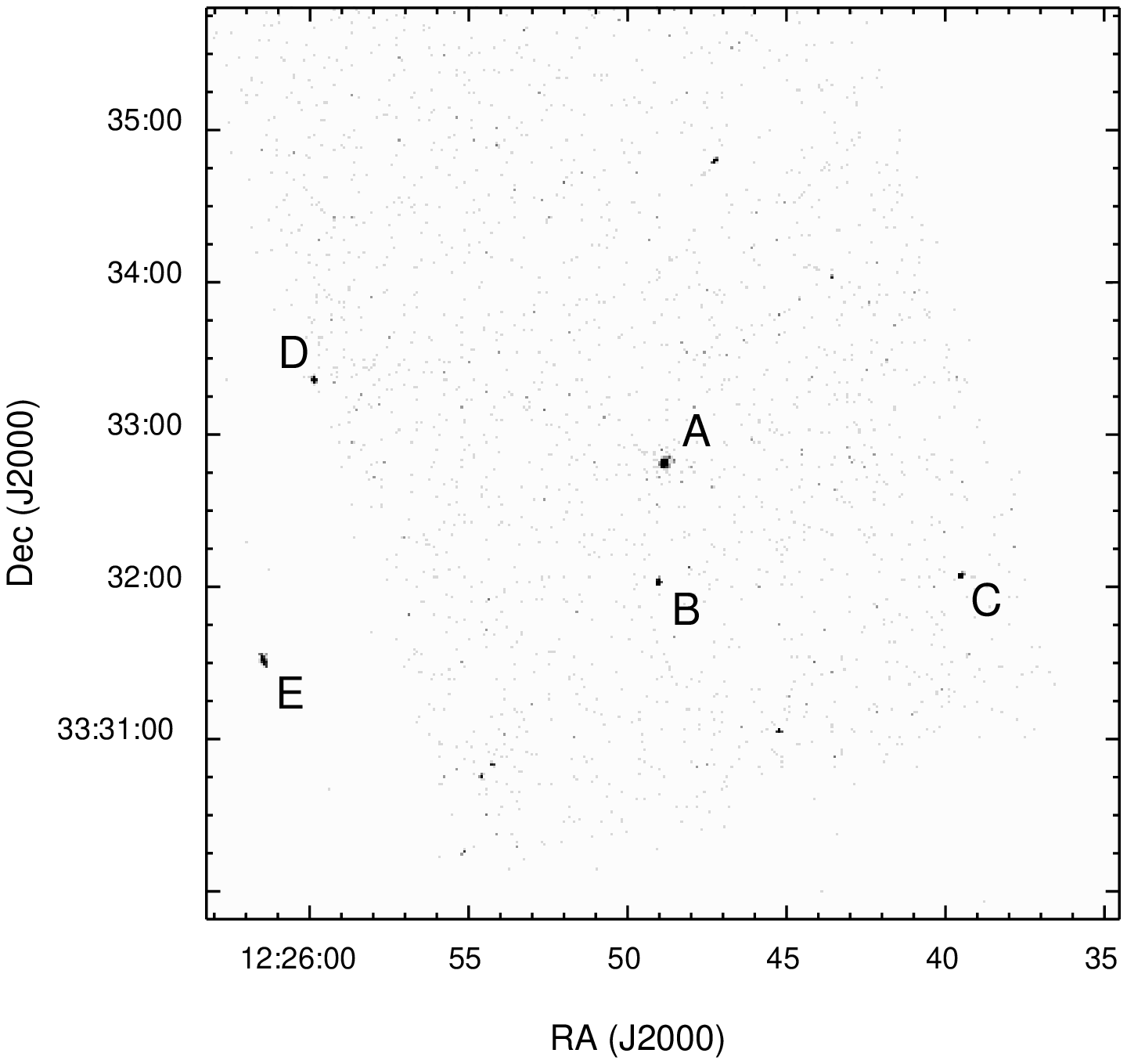,width=8truein,angle=0}}
\vskip 0.2truein
\caption{{\it Chandra\/} ACIS-S image of NGC 4395 in the 0.3--10 keV band.
Sources A--E were previously identified in a {\sl ROSAT\/} PSPC observation
(Moran et al.\ 1999); several other sources are visible as well.  Source~A
is coincident with the nucleus of the galaxy.  It is clear from this image
that the nucleus is completely isolated from the other sources in the
field, which was not the case in the {\sl ASCA\/} observation of NGC~4395
(Iwasawa et al.\ 2000).}
\end{center}
\end{figure}

\clearpage
\begin{figure}
\begin{center}
\centerline{\psfig{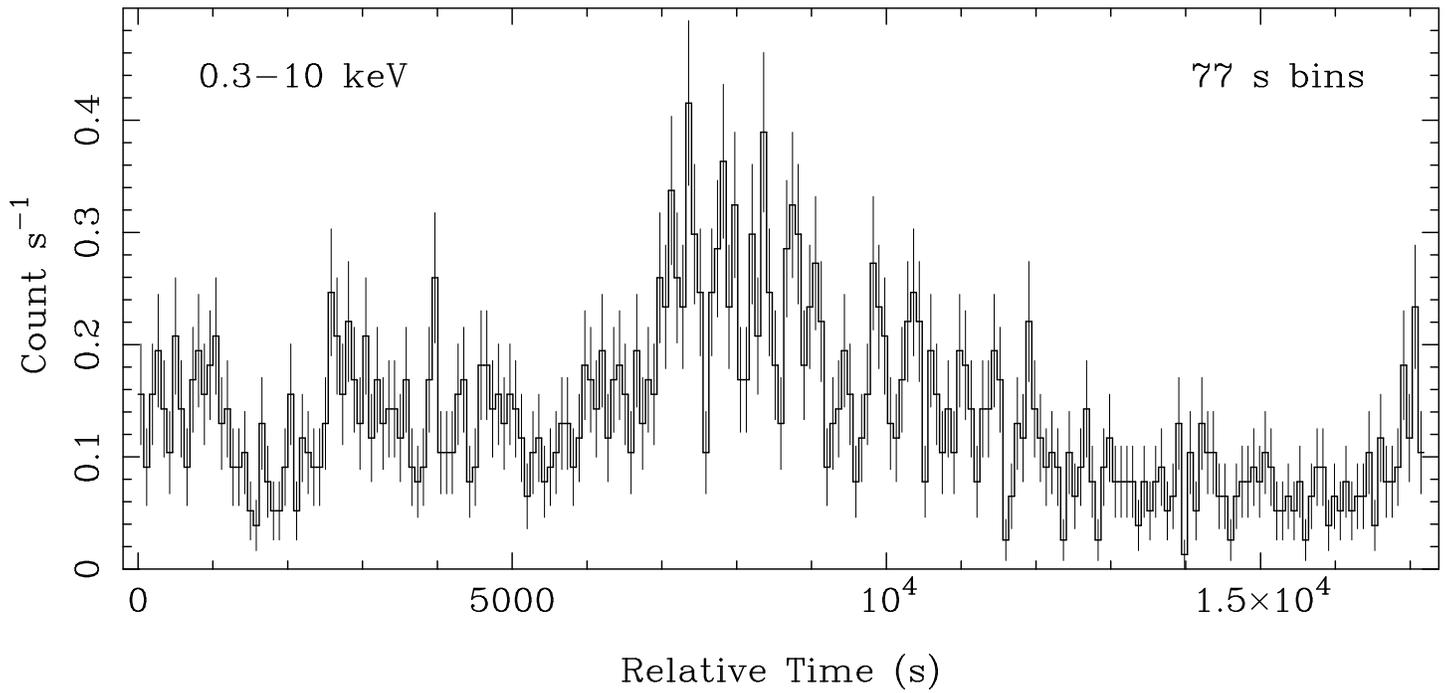}}
\vskip 0.2truein
\caption{The 0.3--10 keV light curve of NGC~4395.  Each 77~s bin 
corresponds to 50 frames.  The source exhibits large-amplitude
variability on very short time scales.}
\end{center}
\end{figure}

\clearpage
\begin{figure}
\begin{center}
\centerline{\psfig{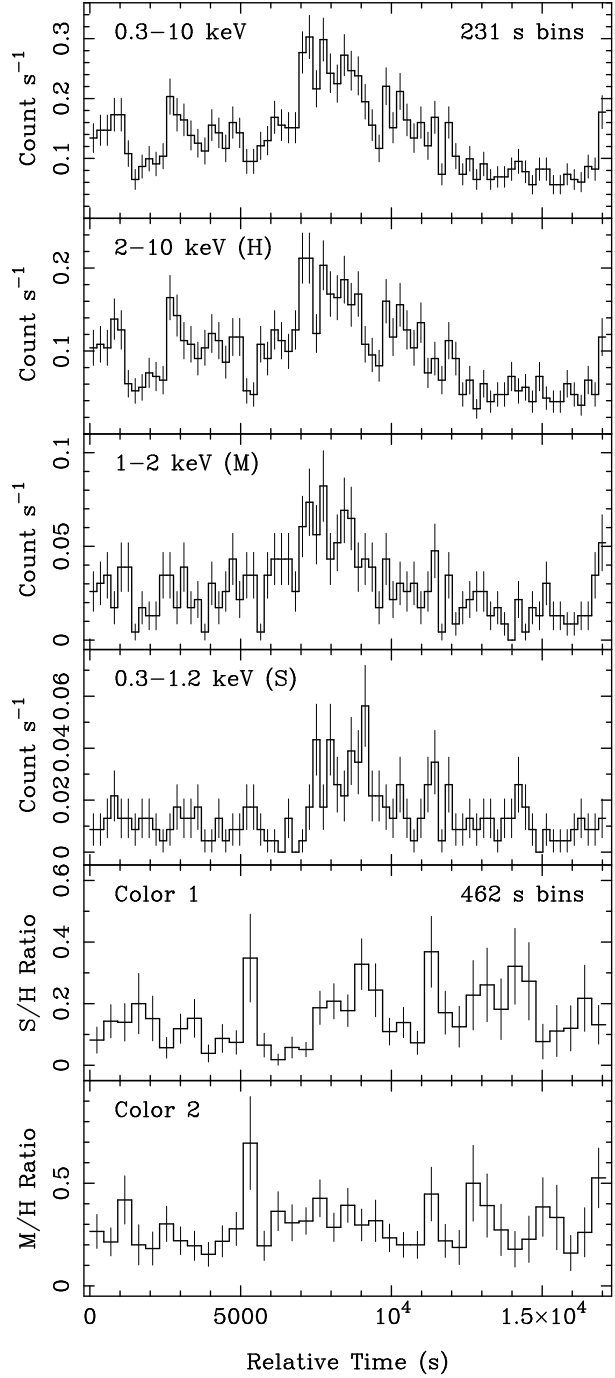}}
\vskip 0.2truein
\caption{Light curves of NGC 4395 in broad, hard (H), medium (M), and
soft (S) energy bands (top four panels).  The 231~s bins correspond
to 150 frames.  The bottom two panels indicate the S/H counts ratio
(Color 1) and the M/H counts ratio (Color 2) for the source in 300-frame
bins.  Note that there is little apparent correspondence between either
color and the 2--10 keV count rate.}
\end{center}
\end{figure}

\clearpage
\begin{figure}
\begin{center}
\centerline{\psfig{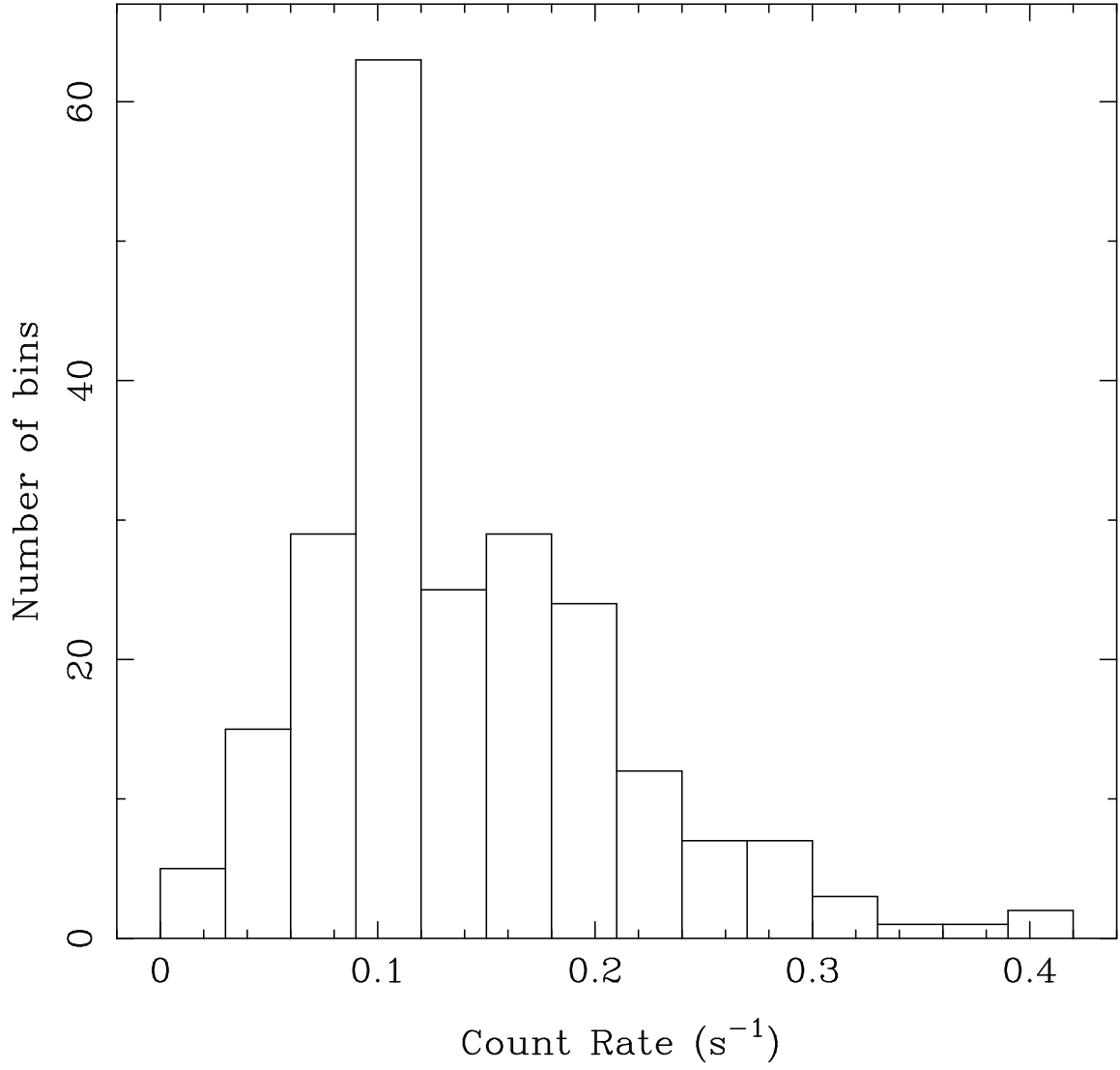}}
\vskip 0.2truein
\caption{Distribution of count rates in the 50-frame light curve shown
in Fig.~2.}
\end{center}
\end{figure}

\clearpage
\begin{figure}
\begin{center}
\centerline{\psfig{figure=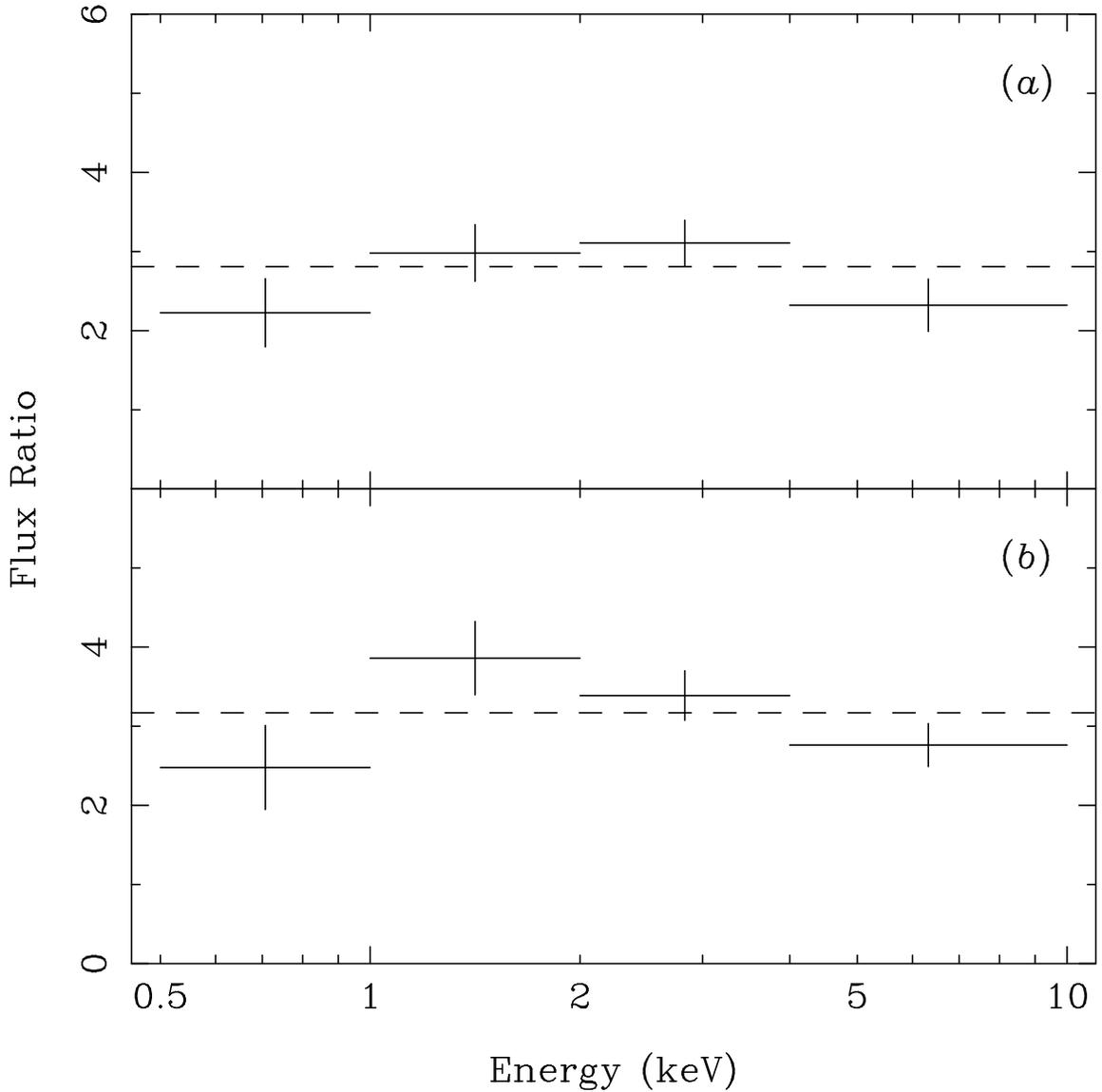,width=6truein,angle=0}}
\vskip 0.2truein
\caption{Ratio of spectra of NGC 4395 in ``active'' and ``quiescent''
states, which have been defined in two different ways (see the text for
details).  In ($a$), the high-state and low-state spectra are associated with
broad periods in the light curve during which the activity of the source
was visibly different.  For ($b$), the high-state and low-state spectra were
derived on the basis of the count-rate histogram shown in Fig.~4.  
The dashed lines denote the mean ratio in the two cases.  Neither plot
indicates strong spectral variability corresponding to the factor-of-three
count rate changes.}
\end{center}
\end{figure}

\clearpage
\begin{figure}
\begin{center}
\centerline{\psfig{figure=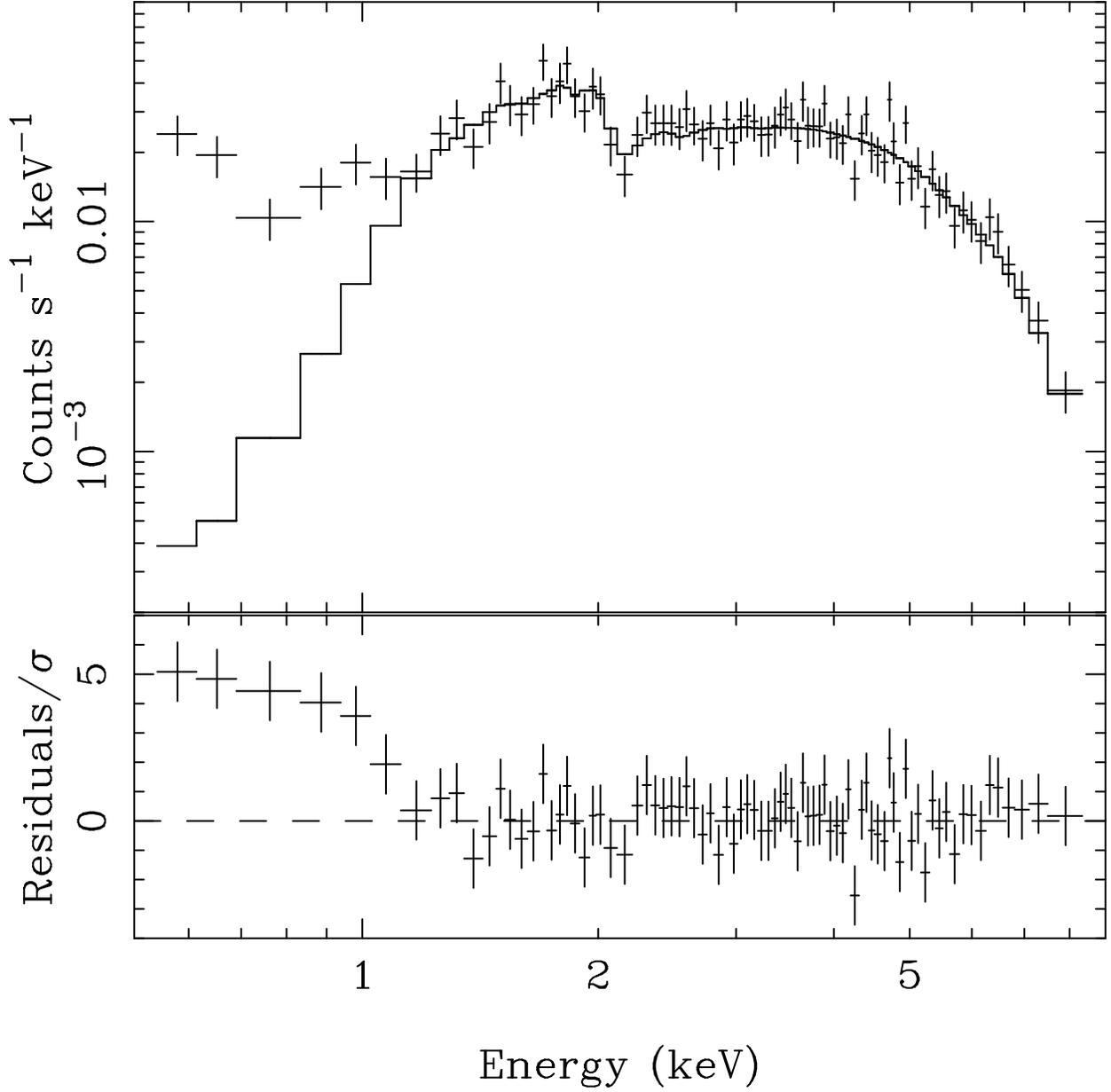,width=6.5truein,angle=0}}
\vskip 0.2truein
\caption{({\it top panel}) Observed ACIS-S spectrum of NGC 4395, fitted with
a simple absorbed power-law model above 1.2 keV ($\Gamma = 0.61, N_{\rm H}
= 1.2 \times 10^{22}$ cm$^{-2}$).  The fit residuals (normalized by the
1$\sigma$ errors) are shown in the lower panel.  Note the strong excess
flux at the lowest energies.}
\end{center}
\end{figure}

\clearpage
\begin{figure}
\begin{center}
\centerline{\psfig{figure=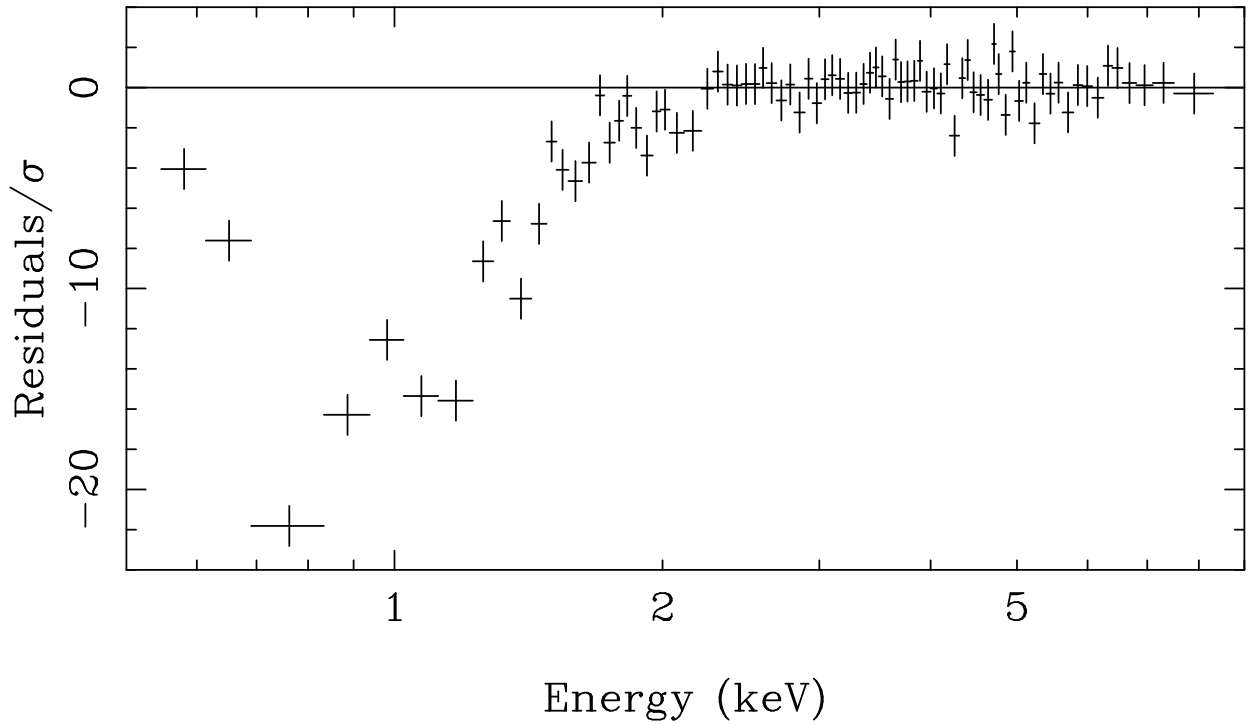,width=6.5truein,angle=0}}
\vskip 0.2truein
\caption{Normalized residuals from a power-law fit to the ACIS-S spectrum
of NGC 4395 above 2.2 keV.  The absorption has been fixed at the Galactic
value.  The deep troughs seen below 2 keV suggest additional absorption
by an ionized medium.}
\end{center}
\end{figure}

\clearpage
\begin{figure}
\begin{center}
\centerline{\psfig{figure=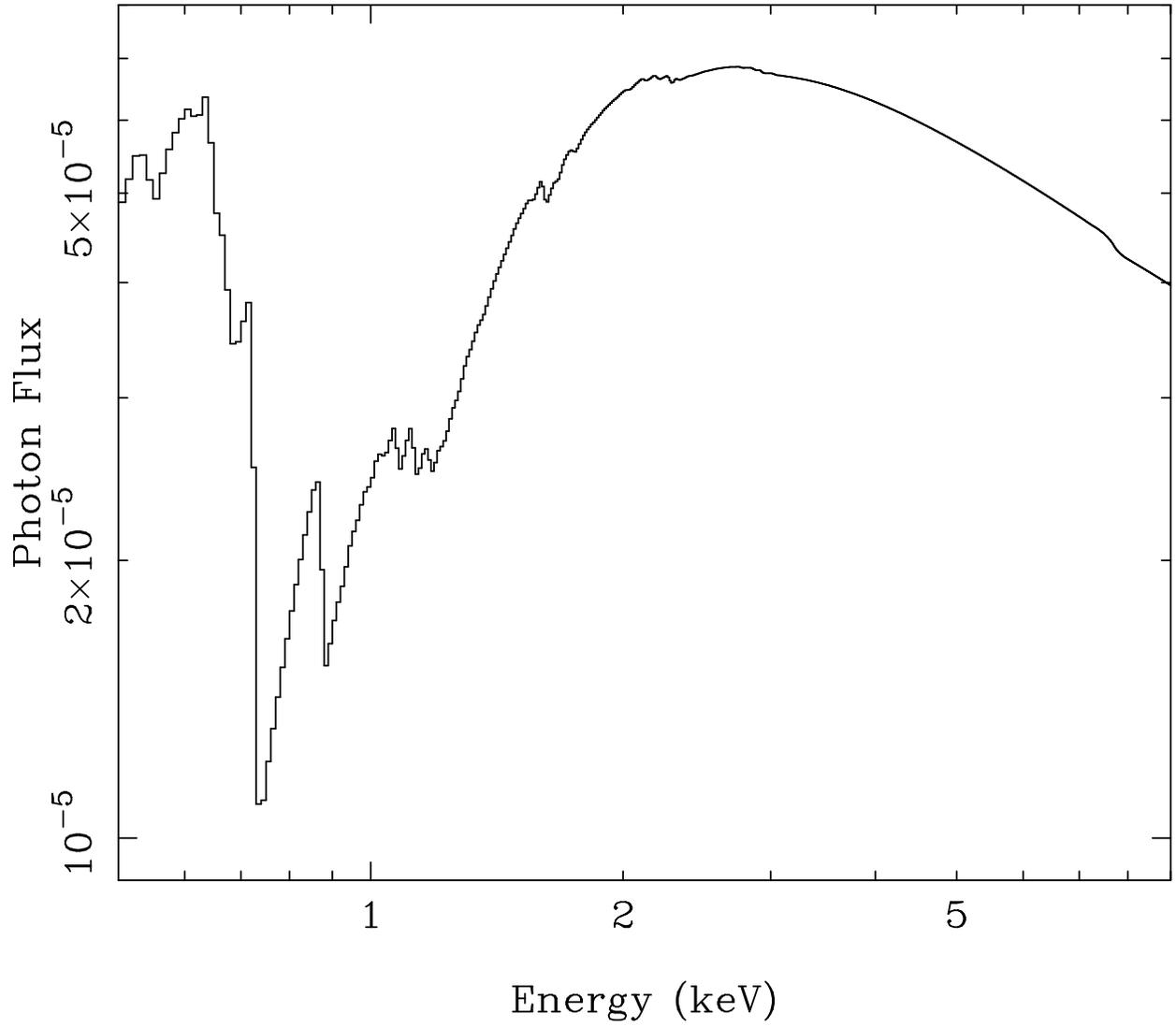,width=6.5truein,angle=0}}
\vskip 0.2truein
\caption{The best-fit single warm absorber model for the ACIS-S spectrum of
NGC~4395.  The ordinate has units of photons cm$^{-2}$ s$^{-1}$
keV$^{-1}$.}
\end{center}
\end{figure}

\clearpage
\begin{figure}
\begin{center}
\centerline{\psfig{figure=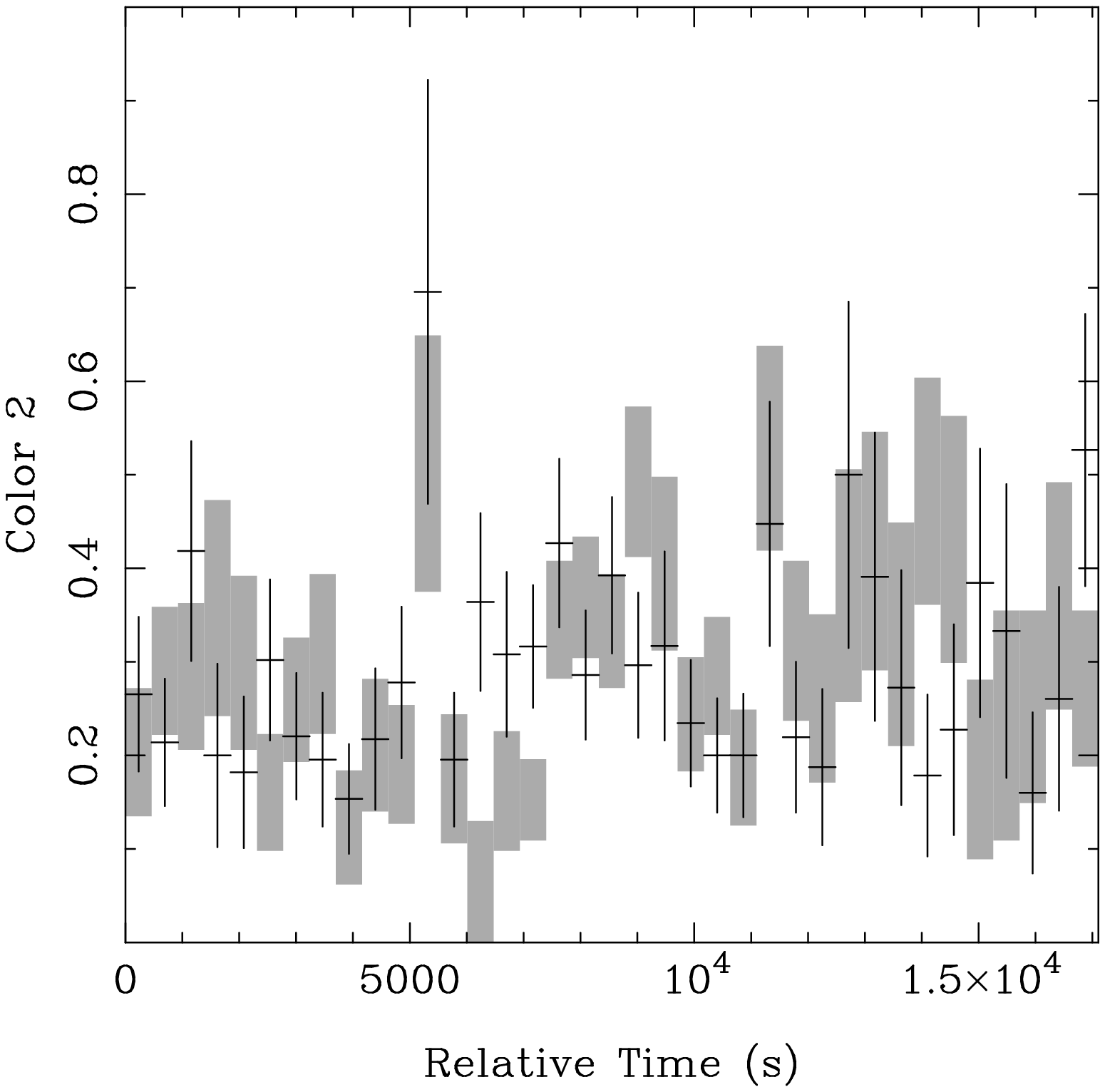,width=6.5truein,angle=0}}
\vskip 0.2truein
\caption{Measured ({\it black crosses}) and predicted ({\it gray bars}) values
of the Color~2 parameter.  The latter were derived from the observed values
of Color~1 shown in Fig.~3, under the assumption that color changes arise
from column density variations in the ionized medium alone.  The length of
the gray bars reflects the uncertainty in Color~1.}
\end{center}
\end{figure}


\begin{thebibliography}{}
\bibitem{}Chartas, G., et al.\ 2000, ApJ, 542, 655
\bibitem{}Elvis, M., Wilkes, B.~J., Giommi, P., \& McDowell, J.\ 1991,
          ApJ, 378, 537
\bibitem{}Eracleous, M., Shields, J.~C., Chartas, G., \& Moran, E.~C.\ 2001,
          ApJ, in press
\bibitem{}Filippenko, A.\ V., \& Ho, L.\ C.\ 2002, ApJ, submitted
\bibitem{}Filippenko, A.\ V., Ho, L.\ C., \& Sargent, W.\ L.\ W.\ 1993,
          ApJ, 410, L75
\bibitem{}Filippenko, A.\ V., \& Sargent, W.\ L.\ W.\ 1989, ApJ, 342, L11
\bibitem{}Halpern, J.~P.\ 1984, ApJ, 281, 90
\bibitem{}Ho, L.~C.\ 1999, ApJ, 516, 752
\bibitem{}Ho, L.~C., et al.\ 2001, ApJ, 549, L51
\bibitem{}Ho, L.~C., \& Ulvestad, J.~S.\ 2001, ApJS, 133, 77
\bibitem{}Iwasawa, K., Fabian, A.~C., Almaini, O., Lira, P., Lawrence, A.,
          \& Hyashida, K.\ 2000, MNRAS, 318, 879
\bibitem{}Kraemer, S.\ B., Ho, L.\ C., Crenshaw, D.\ M., Shields, J.\ C., \&
          Filippenko, A.\ V.\ 1999, ApJ, 520, 564
\bibitem{}Lasota, J.-P., Abramowicz, M.~A., Chen, X., Krolik, J., Narayan, R.,
          \& Yi, I.\ 1996, ApJ, 462, 142
\bibitem{}Lira, P., Lawrence, A., O'Brien, P., Johnson, R.\ A., Terlevich, R.,
          \& Bannister, N.\ 1999, MNRAS, 305, 109
\bibitem{}Matt, G., et al.\ 1996, MNRAS, 281, L69
\bibitem{}Minitti, D., et al.\ 2002, in preparation
\bibitem{}Moran, E.~C., Filippenko, A.~V., Ho, L.~C., Shields, J.~C., 
          Belloni, T., Comastri, A., Snowden, S.~L., \& Sramek, R.~A.\ 1999, 
          PASP, 111, 801
\bibitem{}Murphy, E.\ M., Lockman, F.\ J., Laor, A., \& Elvis, M.\ 1996,
          ApJS, 105, 365
\bibitem{}Nandra, K., George, I.~M., Mushotzky, R.~F., Turner, T.~J., \&
          Yaqoob, T.\ 1997, ApJ, 476, 70
\bibitem{}Nandra, K., \& Pounds, K.~A.\ 1994, MNRAS, 268, 405
\bibitem{}Ptak, A., Yaqoob, T., Mushotzky, R., Serlemitsos, P., \&
          Griffiths, R.\ 1998, ApJ, 501, L37
\bibitem{}Reynolds, C.\ S.\ 1997, MNRAS, 286, 513
\bibitem{}Tucker, W.\ H.\ 1975, Radiation Processes in Astrophysics
          (Cambridge: MIT Press), 168
\bibitem{}Turner, T.~J., George, I.~M., Nandra, K., \& Turcan, D.\ 1999,
          ApJ, 524, 667
\bibitem{}Walter, R,. \& Fink, H.~H.\ 1993, A\&A, 274, 105
\bibitem{}Wrobel, J.~M., Fassnacht, C.~D., \& Ho, L.~C.\ 2001, ApJ, 553, L23
\bibitem{}Zdziarski, A.~A., Leighly, K.~M., Matsuoka, M., Cappi, M., \&
          Mihara, T.\ 2001, ApJ, in press
\end{thebibliography}
\end{document}